\documentclass[fleqn,twoside]{article}
\usepackage{espcrc2}

\usepackage{graphicx}
\newcommand{\AmS}{{\protect\the\textfont2
  A\kern-.1667em\lower.5ex\hbox{M}\kern-.125emS}}
\def\LL{\left\langle}   % left angle bracket
\def\RR{\right\rangle}  % right angle bracket
\def\LP{\left(}         % left parenthesis
\def\RP{\right)}        % right parenthesis
\def\LS{\left[}         % left square bracket
\def\RS{\right]}        % right square bracket
\def\BE{\begin{equation}}
\def\EE{\end{equation}}
\def\BEA{\begin{eqnarray}}
\def\EEA{\end{eqnarray}}
\def\EL{\nonumber\\}

\title{Preliminary results of the heavy-light meson spectrum using chirally 
improved light quarks}

\author{T. Burch\address[RGBG]{Institut f\"ur Theoretische Physik, 
    Universit\"at Regensburg, D-93040 Regensburg, Germany}%
\thanks{Presented by T. Burch.}, 
  C. Gattringer\addressmark\ and 
  A. Sch\"afer\addressmark}

\begin{document}

\begin{abstract}
Using a ``wall'' of quark point sources, we invert the chirally improved
Dirac operator to create an ``incoherent'' collection of quark propagators
which originate from all spatial points of the source time slice.
The lowest-order NRQCD approximation is used to create
heavy-quark propagators from the same wall source.
However, since the numerical cost involved in computing such heavy-quark
propagators is low, we are able to use a number of source gauge paths to
establish coherence between the heavy and light quarks at several spatial
separations.
The resulting collection of heavy-light meson correlators is analyzed to
extract the corresponding mass spectrum.
\vspace{1pc}
\end{abstract}

\maketitle

%In this contribution we present some details and preliminary results of 
%our calculations of the heavy-light meson mass spectrum.

\section{Light-quark source and propagators}

For the light quarks, we use the chirally improved Dirac operator, 
constructed via an expansion in gauge paths and the 
corresponding approximate solution of the Ginsparg-Wilson relation 
(for details see Ref.~\cite{Gins82}).
%\BE
%\gamma_5 D + D \gamma_5 = D \gamma_5 D \, .
%\EE

As sources for the light-quark propagators, we use 12 (3 colors $\times$ 4 
spins) uniform walls, setting the color and spin of the quarks to be the 
same at all spatial positions and relying upon the random gauges of the 
ensemble of configurations to cancel any non-gauge-invariant (open) 
contributions to hadron correlators constructed from these quark propagators. 
This is similar to the volume source used by others \cite{Kura94} 
for disconnected correlators, except for the fact that our source sits on 
only a single time-slice. Connected meson correlators from two of these 
quarks should exhibit a signal arising from $L^3$ closed (or 
gauge-invariant) correlators, along with some additional noise from the 
$L^3(L^3-1)$ open correlators. These sources have been used before for 
heavy quarkonium correlators with some success \cite{Burc01} since the 
signal-to-noise ratio proved to be surprisingly good. Since we retain 
our light-quark propagators and the heavy-quark evolution is 
computationally cheap, we are free to use a number of different gauge paths 
to create ``walls'' of extended heavy-light meson sources from the 
light-quark source before we compute the heavy-quark propagators.
In this way we can construct a full matrix of correlators with 
various spatial extents and a collection of starting positions without 
additional light-quark inversions.

\section{Heavy-quark evolution}

The heavy-quark propagators are created using lowest-order NRQCD, along 
with some improvements for the lattice version \cite{Lepa92}. 
A single time-step of heavy-quark evolution is achieved via
%(suppressing color indices for the moment)
\BEA
\label{HQE}
G_Q(\vec x, 0 | \vec y,t+a) = \LP 1 - \frac{a\tilde H_0}{2n}\RP^n_{t+a}
U^\dagger_4(\vec y\,'',t) \EL
\LP 1 - \frac{a\tilde H_0}{2n}\RP^n_t G_Q(\vec x, 0 | \vec y\,', t) \, ,
\EEA
where we use $n=2$ (this is sufficient for $b$ quarks on our lattices) and 
the Hamiltonian is the $O(a^2)$-improved, covariant, kinetic-energy operator,
corrected to remove $O(a)$ artifacts in the time evolution:
\BE
\tilde H_0 = - \frac{\tilde\Delta^2}{2m_Q^{}} - 
\frac{a\tilde\Delta^4}{16nm_Q^2} \, .
\EE
Since we include only the kinetic-energy term in our NRQCD expansion, there 
is no coupling of the heavy quark's color and spin, the latter being 
provided simply by the projection
\BE
G_{\alpha\beta;Q}^{ab}(\vec x, 0 | \vec y,t) = \frac{1}{2}
(1 + \gamma_4)_{\alpha\beta} G_Q^{ab}(\vec x, 0 | \vec y,t) \, .
\EE

\section{Source and sink operators}

The gauge paths and corresponding spin structures we use to create our 
mesons can be found in Refs.~\cite{Laco96,Mich98}. Through sums ($s$) and 
differences ($p$) of paths in opposite directions, along with different 
Dirac matrices, meson operators with different symmetries can be constructed: 
e.g.,
\BE
\label{meson_S}
S : \overline Q(x) \Gamma U_P(x|x') q(x') = 
\overline Q \, \gamma_5 \sum_i s_i \, q \, ;
\EE
\BE
P_- : \overline Q \, \sum_i \gamma_i p_i \, q
\; ; \;
P_+ : \overline Q \LP \gamma_1 p_1 - \gamma_2 p_2 \RP q \, ;
\EE
\BE
\label{meson_D}
D_\pm : \overline Q \, \gamma_5 \LP s_1 - s_2 \RP q \, .
\EE
Thus far, we have only included straight paths, limiting us to S-, P- and 
D-wave mesons. 
Also, since we don't include heavy-quark spin interactions, a number of 
physical states are not separable: the S-wave operator, $S$, must be 
viewed as creating a spin-average of $J^P=0^-$ and $1^-$ states; 
the $P_-$, $J^P=\{0^+,1^+\}$; the $P_+$, $J^P=\{1^+,2^+\}$; and 
the $D_\pm$, $J^P=\{1^-,2^-,3^-\}$.

\section{Heavy-light correlator matrix}

Given all the ingredients, we now present the form of our heavy-light 
correlators:
\BEA
C_{ij}(t) \!\!\!\!\!&&\!\!\!\!\! = \LL
\sum_{\vec x,\vec x\,',\vec y}
\LS \Gamma_{\delta\alpha} U_P^{da}(x'|x'') \RS_i
G_{\alpha\beta;q}^{ab}(\vec x,0|\vec y,t)
\right. \EL && \left.
\LS \Gamma'^*_{\beta\gamma} U_{P'}^{*bc}(y|y') \RS_j
G_{\gamma\delta;\overline Q}^{cd}(\vec y\,',t|\vec x\,',0)
\RR .
\EEA
Note the sum over both quark source positions, $\vec x$ and $\vec x\,'$, 
due to the use of the walls; only the closed terms (where 
$\vec x = \vec x\,''$) will contribute to the signal. 
We include both the $\overline Q q$ and $Q \overline q$ 
correlators from opposite time directions.

After calculating the matrix of correlators, $C_{ij}(t)$, between our 
linearly independent meson operators, 
$\psi_i = \overline Q(x) \Gamma U_P(x|x \! + \! r_i) q(x \! + \! r_i)$, we 
solve the corresponding generalized eigenvalue problem to extract the 
physical states \cite{Mich85}:
\BE
\sum_j C_{ij}(t) \psi_j^k = \lambda^k(t,t_0) \sum_j C_{ij}(t_0) \psi_j^k \, ,
\EE
where, for large enough values of $t$ ($\gg t_0$),
\BE
\lambda^k(t,t_0) = e^{-m_k (t-t_0)} \, .
\EE
So the eigenvalues provide the masses of the states and the eigenvectors 
reveal the meson wavefunctions' dependencies upon the $\overline Q q$ 
separation, $r_i$ ($=0-7a$ along the principal axes). This technique of 
using operators with different spatial extents has also been recently used 
to determine excited states of baryons \cite{Burc04}.

\section{Preliminary results}

In this section we present results from one set (100) of $12^3\times24$ 
quenched configurations which were created using the L\"uscher-Weisz action 
at $\beta=7.9$ ($r_0/a=3.384$ $\rightarrow$ $a^{-1} \approx 1330$ MeV).

\begin{figure}[t]
\resizebox{2.9in}{!}{\includegraphics{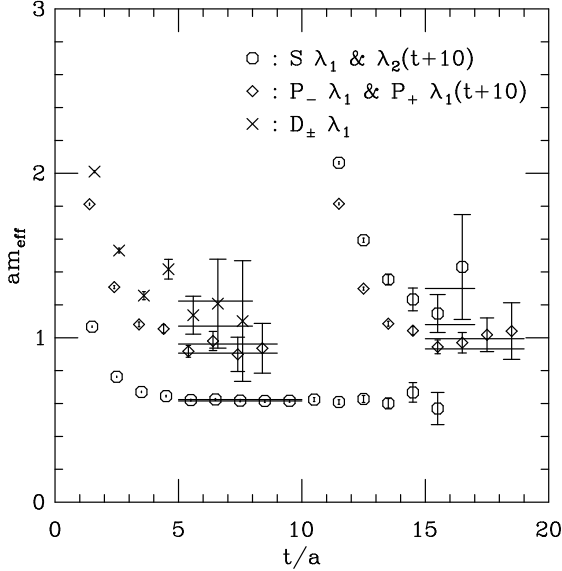}}
\caption{
\label{effmass}
Effective masses versus time from the largest eigenvalues of each of the 
meson operators. 
The results for the $P_+$ and for the first-excited state of the $S$ are 
shifted by 10 in time for clarity. 
$am_q^{}=0.08$, $am_Q^{}=3.0$ and $t_0/a=1$.
}
\end{figure}

In Fig.~\ref{effmass} we plot the effective masses calculated from the 
largest eigenvalues of each set of operators. The horizontal lines indicate 
the $1\sigma$ masses resulting from correlated fits to the data over the 
corresponding time ranges. We repeat this for 6 mass combinations 
($am_Q^{}=3.0,3.5$ and $am_q^{}=0.04,0.08,0.10$) in order to interpolate to 
the physical values of $m_b^{}$ and $m_s^{}$, and to extrapolate to 
$m_{u(d)}$. The physical strange quark mass is determined via light-light 
meson spectroscopy ($M_{s\bar s}^2 \approx 2M_K^2 - M_\pi^2$ $\rightarrow$ 
$am_s^{}\approx 0.09$); the bottom quark mass is determined via 
finite-momentum heavy-light states and the corresponding dispersion relation:
\BE
E(p) \approx E_0 + \frac{\vec p\,^2}{2M} \, ,
\EE
where we require that $M=(3M_{B_s^*}+M_{B_s})/4$. 
At this lattice spacing we find $m_b^{} \approx 4240$ MeV. 
The interpolated (and extrapolated) results for the $B_s^{(*)}$ ($B^{(*)}$) 
system are displayed in Table \ref{Bs_splittings} (Table \ref{B_splittings}).

\begin{table}[ht]
\caption{\label{Bs_splittings}
Mass splittings for the $B_s^{(*)}$ system.}
\begin{tabular}{llll} \hline
${\cal O}$ & $a(M-M_S)$ & $M-M_S$ (MeV) & exp.~\cite{Eide04} \\ \hline
$S\,'$ & 0.58(11) & 770(150)(?) & - \\
$P_-$ & 0.314(28) & 419(37) & 448(16) \\
$P_+$ & 0.341(31) & 455(41) & 448(16) \\
$D_\pm$ & 0.530(76) & 710(100) & - \\ \hline
\end{tabular}
\end{table}

\begin{table}[ht]
\caption{\label{B_splittings}
Mass splittings for the $B^{(*)}$ system.}
\begin{tabular}{llll} \hline
${\cal O}$ & $a(M-M_S)$ & $M-M_S$ (MeV) & exp.~\cite{Eide04} \\ \hline
$S\,'$ & 0.46(17) & 610(230)(?) & - \\
$P_-$ & 0.306(50) & 408(67) & 384(8) \\
$P_+$ & 0.330(58) & 440(77) & 384(8) \\
$D_\pm$ & 0.45(11) & 600(150) & - \\ \hline
\end{tabular}
\end{table}

The $SP$ and $SD$ mass splittings are in agreement with experiment 
(available only for the former) and the findings of previous lattice studies 
\cite{Mich98,Alik00,Gree04}. 
All ground-state eigenvector components have the same sign: i.e., there are 
no nodes in the corresponding ground-state heavy-light wavefunctions.
The first-excited S-wave displays a single node at 
$r \approx 2a \approx 0.6r_0$ ($r \approx 0.5r_0$ found in \cite{Mich98}).
Our mass results for the first-excited S-wave, 
however, appear to be higher than those found previously 
($\approx 500$ MeV for $B_s^{}$). Our mass plateaus for this state are rather 
suspect (typically only 3 $t$ values were used in the fits). This issue can 
only be resolved with better statistics, which can be achieved a number of 
ways: more or larger configurations, more wall sources with locally 
gauge-rotated quark fields, more operators (e.g., non-straight paths). 
We would like to consider each of these options in the future, along with the 
inclusion of heavy-quark spin interactions to resolve more spin splittings in 
these heavy-light systems.

This work is supported by BMBF and GSI. All computations were performed 
on the Hitachi SR8000 in Munich and at the RZ Regensburg.

\end{document}